**Achromatic metasurfaces with inversely customized dispersion for ultra-broadband acoustic beam engineering**

**Authors**
Hao-Wen Dong[1,2,3]†, Chen Shen[4,5]†, Sheng-Dong Zhao[6]†, Weibao Qiu[7]†, Juan Zhou[7], Chuanzeng Zhang[8], Hairong Zheng[7], Steven A. Cummer[4], Yue-Sheng Wang[1]*, Li Cheng[2]*

**Affiliations**
[1]Deparment of Mechanics, School of Mechanical Engineering, Tianjin University, Tianjin 300350, PR China.
[2]Department of Mechanical Engineering, The Hong Kong Polytechnic University, Hong Kong, PR China.
[3]Department of Applied Mechanics, University of Science and Technology Beijing, Beijing 100083, PR China.
[4]Department of Electrical and Computer Engineering, Duke University, Durham, North Carolina 27708, USA
[5]Department of Mechanical Engineering, Rowan University, Glassboro, NJ, 08028, USA
[6]School of Mathematics and Statistics, Qingdao University, Qingdao 266071, PR China
[7]Paul C. Lauterbur Research Center for Biomedical Imaging, Institute of Biomedical and Health Engineering, Shenzhen Institutes of Advanced Technology, Chinese Academy of Sciences, Shenzhen 518055, PR China
[8]Department of Civil Engineering, University of Siegen, D-57068 Siegen, Germany

*Correspondence to: yswang@tju.edu.cn (Y. S. Wang); li.cheng@polyu.edu.hk (L. Cheng)
†These authors contributed equally to this work.

**Abstract**

Metasurfaces, the ultrathin media with extraordinary wavefront modulation ability, have shown versatile potential in manipulating waves. However, existing acoustic metasurfaces are limited by their narrow-band frequency-dependent capability, which severely hinders their real-world applications that usually require customized dispersion. To address this bottlenecking challenge, we report ultra-broadband achromatic metasurfaces that are capable of delivering arbitrary and frequency-independent wave properties by bottom-up topology optimization. We successively demonstrate three ultra-broadband functionalities, including acoustic beam steering, focusing and levitation, featuring record-breaking relative bandwidths of 93.3%, 120% and 118.9%, respectively. All metasurface elements show novel asymmetric geometries containing multiple scatters, curved air channels and local cavities. Moreover, we reveal that the inversely designed metasurfaces can support integrated internal resonances, bi-anisotropy and multiple scattering, which collectively form the mechanism underpinning the ultra-broadband customized dispersion. Our study opens new horizons for ultra-broadband high-efficiency achromatic functional devices on demand, with promising extension to the optical and elastic achromatic metamaterials.

**Introduction**

Metasurfaces have shown tremendous possibility in the fields of optical devices (*1-9*), carpet cloaking (*9*), medical ultrasound (*10*, *11*), architectural acoustics (*12*) and nondestructive testing (*13*). Acoustic metasurfaces (AMs) (*10-12*, *14-24*), as an important category of metamaterials, have revolutionized the way for controlling the absorption, reflection and transmission of acoustic waves due to their extraordinary wavefront-shaping ability. Examples include near-perfect sound absorber (*15*), noise control (*14*, *22*), Schroeder diffuser (*12*), cloaking (*16*), beam steering (*10*, *11*, *20*, *23*), focusing (*18*, *23*, *24*), asymmetric transmission (*19*), vortex beam (*17*) and acoustic levitation (*11*). Irrespective of the specific functionality, the kernel of the metasurface design is to construct elements that possess unique refractive index and impedance to control the amplitude and the phase shift of transmitted or reflected waves. Existing airborne sound metasurfaces are mostly based on Helmholtz-resonator (*12*, *16-20*, *23*), space-coiling (*10*, *11*, *14*, *24*) or membrane-type (*15*) elements. While each of these configurations has its unique features and advantages, there is a lack of a unified and generic design approach for the realization of AM elements that exhibit desired universal performance. For example, space-coiling structures can generate a large range of phase delay by properly folding the wave propagation path. The transmission, however, is usually compromised and high transmission typically only occurs within a narrow band. The lack of freedom in the design space hampers the realization of novel wave functionalities. More importantly, almost all AMs suffer from narrow bandwidth, leaving their real-world applications almost untapped. Although coding (*18*) or tunable (*24*) approaches can, in principle, expand the bandwidth by adjusting the element configuration, they undoubtedly increase the system complexity and cost, thus making the devices impractical. Besides, with multiple-frequency waves simultaneously impinging upon a metasurface, it is difficult, if not impossible, to achieve the same wave functionality only by tunable approaches. Therefore, achieving broadband frequency-independent properties calls for the conception of new AM architectures beyond the existing wave mechanisms, an unavoidable step toward the development of practical acoustic functional devices.

Topology optimization, a dedicated inverse-design methodology, has been successively employed to achieve high-performance wave devices in electromagnetics (*25-27*), acoustics (*28-29*) and elastics (*30*), leading to the new topological discoveries, wave mechanisms and even broadband properties. In recent two years, the machining learning was also introduced to perform the metamaterial engineering (*31*). However, most existing studies focused on various wave functionalities and phenomena of metamaterials with the inevitable strong dispersions. From the application perspective, the ultra-broadband frequency-independent feature, with relative bandwidths larger than 100%, is still lacking up to now, especially for metasurfaces. Capitalizing on its unique ability in simultaneously handling geometrical and physical properties of a structure in a large search space, an inverse-design methodology might offer the ideal tool for conceiving novel metasurface elements with distinct features.

To achieve an arbitrary and broadband functionality, each metasurface element has to deliver required local amplitude and/or phase modulation that are controlled across the entire operating bandwidth. Such modulation depends on the specific functionality of the metasurface and translates to customized dispersion of each element. Recent work has shown that broadband (*4-7*) optical multi-wavelength achromatic (*32*) devices can be achievable by using sophisticated artificial structures (*4-7*, *32*). Their broadband performance, however, still relies on achromatic phase modulation of the elements (*4-7, 32*), rather than completely arbitrary dispersion considering the constituent elements and whole metasurface macrostructure. This possibly leaves a huge potential space to reach the limit broadband and even ultra-broadband achromatic properties of metasurfaces. Consequently, in the whole fields of both optical and acoustic metasurfaces, the most fundamental and crucial challenge is how to systematically construct the broadband and even ultra-broadband metasurfaces with arbitrary desired microscopic and macroscopic dispersions simultaneously.

To cope with the aforementioned challenges, in this Article, we propose a systematic bottom-up inverse design for the realization of ultra-broadband and customized metasurfaces and explore the underlying physics underpinning different functionalities. Our study demonstrates the feasibility of developing a systematic inverse-design model for constructing ultra-broadband achromatic metasurfaces whose dispersions can be purposely tailored. The optimized elements can either be dispersionless or exhibit customized dispersions, showing multiple asymmetric scatters and cavities within the element space. By combining these elements in specific patterns, we successively construct inversely designed metasurfaces, which entail ultra-broadband, namely achromatic beam steering, focusing and levitation, beyond the relative bandwidths of all existing achromatic metasurfaces. Finally, we also reveal the ultra-broadband mechanism which emerges from a combination of integrated internal resonances, bi-anisotropy and multiple scattering effects. We construct three categories of brand-new ultra-broadband achromatic metasurfaces with customized microscopic and macroscopic dispersions, which entail record-breaking ultra-broadband frequency-independent acoustic beam functionalities, never achieved before. The uncovered asymmetric and bi-anisotropic microstructural topological feature and the synthetical ultra-broadband achromatic mechanism offer a new basic platform for designing broadband frequency-independent metasurfaces. Meanwhile, the inversely designed metasurface gives rise to an ultra-broadband, stable and single-sided ultrasound levitation, which can hardly be accomplished by the traditional ultrasound levitation technologies. The present customized ultra-broadband AMs are expected to enable applications in medical ultrasonics and contactless particle control assembly. Meanwhile, the proposed ultra-broadband achromatic strategy can be extended to optical and elastic metasurfaces to achieve high-efficiency frequency-independent advanced functional devices.

**Results**

Figure 1 illustrates the schematic of ultra-broadband metasurfaces consisting of inversely designed elements. When waves impinge upon the metasurface, the frequency-independent functionality will be accomplished within a broad frequency range if all output wavefronts are identical at different frequencies (Fig. 1A). For example, this typical achromatic feature will generate a fantastic bottle beam with nearly the same focusing location at every operating frequency. Its particular acoustic amplitude isosurface can further provide the acoustic radiation pushing and pulling forces to levitate a particle in the mid-air within the ultra-broadband range. Such an ultra-broadband levitation phenomenon can find promising scientific and industrial applications in the fields of the contactless particle/droplet/biological cell control and assembly (33), even a volumetric display technique (33). Of course, the elaborately inversely designed ultra-broadband achromatic metasrufaecs can lead to ultra-broadband customized achromatic functionalities such as anomalous beam steering (10), focusing (10, 11) and even levitation (11) (Fig. 1B) with the beam characteristics on demand. In principle, broadband wavefront manipulation is tied to specific dispersion of every element and meticulous combination of all elements (Fig. 1C), i.e., the microscopic and macroscopic dispersions of metasurfaces. For beam steering, every element should maintain a constant effective index within the frequency range. As the required phase shift increases linearly between adjacent elements, the index across all the elements also increases linearly at the same frequency. Similarly, for focusing, every element has to maintain a constant index. The difference between the indices of adjacent elements, on the other hand, varies according to element position so that the same focal point can be achieved. However, for levitation, the elements must support specific dispersion (herein referred to as customized dispersion), i.e., the indices are kept constant for some elements of the metasurface while decrease nonlinearly with the increase of frequency for others. The variation of the indices among different elements differs from the other two functionalities. Obviously, these complex and rigorous dispersions, alongside the complex phase shift distributions of elements (Fig. 1D), make the broadband design extremely challenging.

Here, we employ the bottom-up topology optimization (Method M1, Supplementary text S1) based on a genetic algorithm (GA) (*28*-*30*) to generate the required phase shift of every element within the ultra-broadband frequency range of interest. Through evolution, the algorithm can produce optimized elements in binary pixels (Fig. 1E), whose assembly gives the final metasurface which ensures the desired functionalities. Different from the top-down topology optimization, our microstructure-macrostructure-functionality design paradigm has the following advantages: i) avoiding the larger-scale computing of functionalities; ii) reducing the search space for lower optimization complexity; iii) providing rich customized topologies for the unit library; and iv) conducive to the exploration of topological features and mechanisms. Note that our elaborately proposed optimization problem of the ultra-broadband metasrufaces with the customized dispersions can be effectively solved by the other evolutionary algorithms, gradient algorithms, and even the machine learning as well.

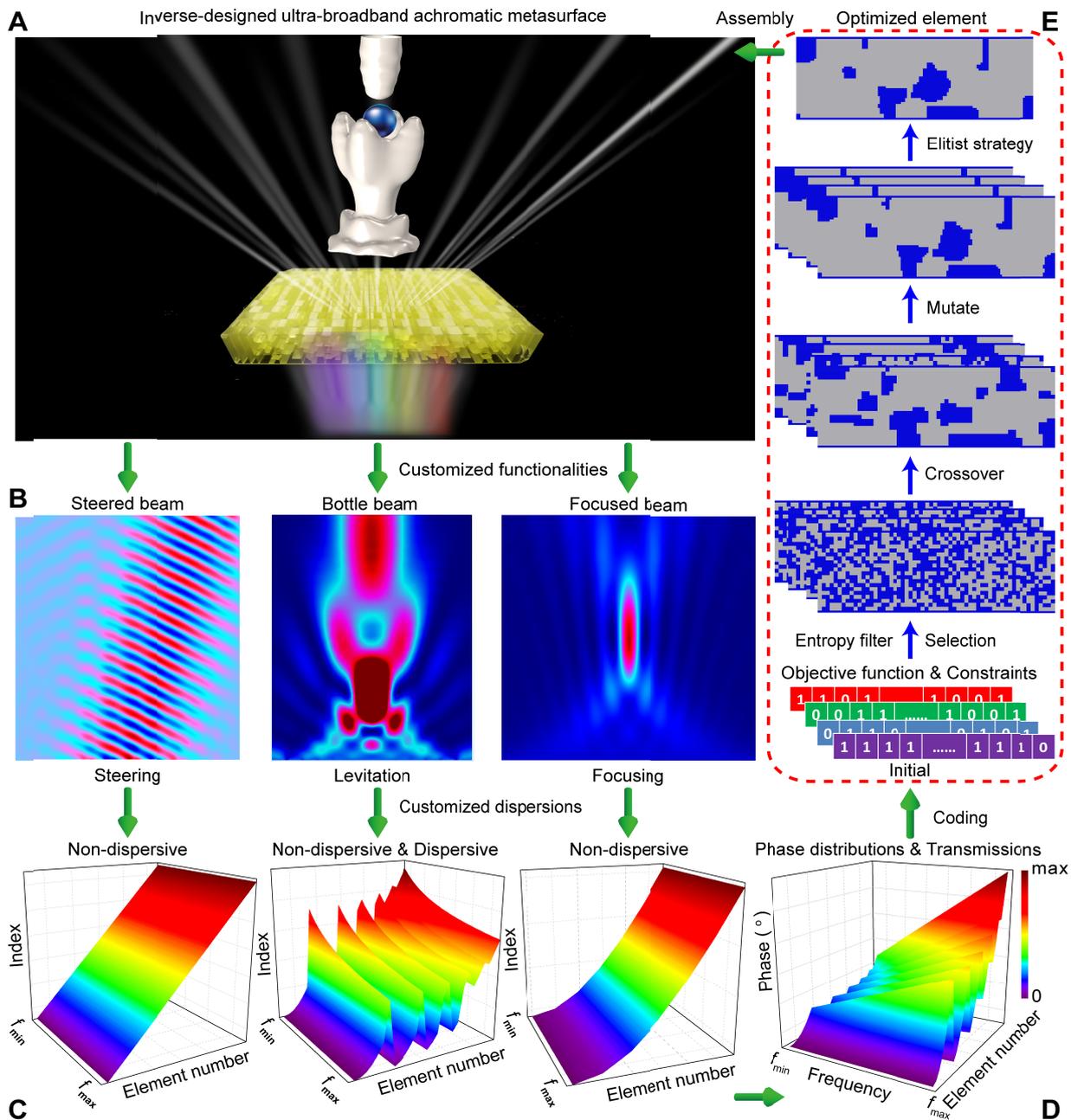

**Fig. 1. Schematic of ultra-broadband achromatic metasurfaces designed by bottom-up topology optimization.** (**A**) An inversely designed ultra-broadband achromatic metasurface (the middle yellow block) which can transform the multi-wavelength incident acoustic waves into an exotic bottle beam with nearly the same acoustic amplitude isosurface (two white stereoscopic objects) at different frequencies within [$f_{min}$, $f_{max}$], making a particle suspended in the same location of the mid-air. Different colors under the metsaurface represent different operating frequencies/wavelengths. (**B**) Anomalous beam steering, focusing and levitation by the inverse-designed metasurface. (**C**) Required effective indices of all metasurface elements for beam steering (left), focusing (right) and levitation (middle) in (B). (**D**) Target phase distributions of elements for a customized functionality. (**E**) Sketch of the heuristic bottom-up topology optimization (Method M1, Supplementary text S1). Under the drive of objective function and constraints, the binary algorithm starts from a randomly initial population and gradually generates a customized optimized element through a series of genetic operations and structural entropy filters. Grey and blue pixels represent the air and solid media, respectively. All optimized elements are assembled into an inversely designed metasurface for a customized functionality.

## Engineering metasurface for ultra-broadband achromatic beam steering

We begin with anomalous refraction (*10*) to validate the proposed inverse-design strategy. To achieve broadband negative refraction with a fixed angle of 20.2°, we perform topology optimization (Methods M1 and M2) of 2D metasurface elements (Supplementary text S3, Supplementary movies S6-S8) based on generalized Snell's law (*1*). As illustrated in Fig. 2A, except Element #1, all the optimized elements have common topological features: i) highly asymmetric geometries; ii) complex curved air channels with nonuniform widths; and iii) distributed cavities located in several local domains. In particular, Elements #5 and #6 have similar shapes while Elements #2, #3 and #4 are different. This indicates that broadband non-dispersive properties can be achieved through an optimal combination of topologies rather than a single configuration. Previous design strategies using only Helmholtz-resonator or space-coiling topology (*10-12*, *14*, *16-20*, *23*, *24*) can hardly deliver such broadband functionality. The topologically optimized structures possess ultra-broadband constant effective indices (Fig. 2B). Note that the final operation range becomes [1600 Hz, 4400 Hz] although the target frequency range is [2000 Hz, 4000 Hz] in the optimization (Method M1). This distinct broadening effect implies the relatively stable performance. The transmission constraint (Method M1) also ensures that all elements offer a transmission coefficient exceeding 80% within almost the entire frequency band (Fig. 2C). The achieved high-transmission performance is mainly attributed to the presence of dominant air-borne sound transmission paths in all optimized elements. This phenomenon is widely seen in classic sound transmission problems on a compound surface with air leakage. It is noted that, the optimized elements do not exhibit the Fabry-Perot (FP) resonance effect whenever their transmissions are relatively high (Supplementary Fig. S7b). Conversely, the FP resonance effect occurs within a very narrow frequency range only in the case of apparent low transmission (Supplementary Figs. S13b and 20c). As a result, the assembled inversely designed metasurface clearly gives rise to ultra-broadband anomalous refraction with requested constant angle (Figs. 2D-2E), even when thermo-viscous loss (*14*, *34-36*) is considered (Supplementary text S4). Since the supercell (i.e., Elements #1-#6) cannot ensure a full $2\pi$ phase coverage at all frequencies, the inversely designed metasurface inevitably incurs phase discontinuity at the interface of two periods, causing some near-field scattering (Supplementary text S4). Nevertheless, the optimized asymmetric element provides a new platform for ultra-broadband beam steering.

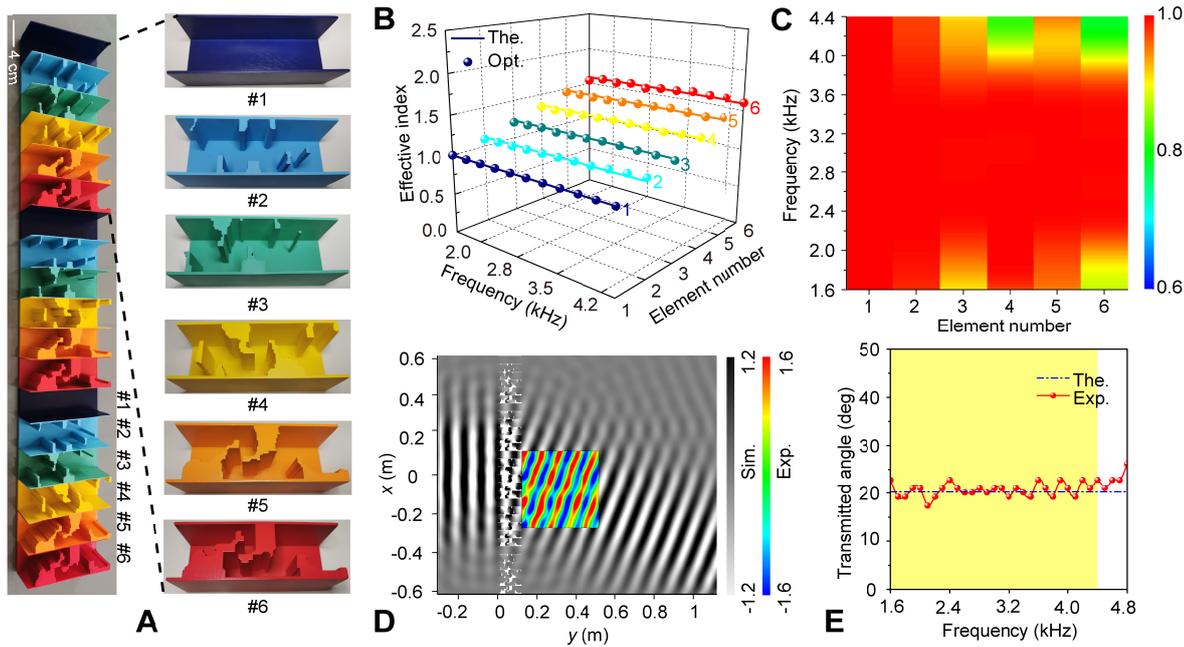

**Fig. 2. Ultra-broadband achromatic beam steering using and inversely designed metasurface.** (**A**) A fabricated gradient metasurface with 3 periods each containing 6 optimized elements. (**B-C**) Theoretical (The.), optimized (Opt.) effective indices (**B**) and transmission coefficients (**C**) of all optimized elements within [1600 Hz, 4400 Hz]. (**D**) Simulated acoustic pressure field at 4000 Hz. The inset shows the measured acoustic field, which is in good agreement with the numerical results. (**E**) Measured transmitted angle through the inversely designed metasurface. The dash dot line denotes the theoretical desired refraction angle 20.2° which is arbitrarily prescribed. The yellow shadow area represents the bandwidth of the metasurface within [1600 Hz, 4400 Hz].

**Engineering metasurfaces for ultra-broadband achromatic focusing**

Wave focusing is a fundamental task pertinent to a wide range of applications including novel integrated optics (*3*, *5*, *6-8*) and ultrasonic imaging (*10*, *11*). Unlike optical focusing (*4-7*), frequency-independent acoustic focusing has not been demonstrated. Here we use the aforementioned inverse-design strategy (Methods M1 and M3) to systematically construct a metasurface for ultra-broadband focusing at a constant focal depth (Fig. 3A). The optimized elements have similar topological features with those for ultra-broadband beam steering (Supplementary text S5). Intriguingly, Elements #4, #5, #6 and #7 have similar overall topologies, i.e., four solid blocks asymmetrically distributed in the domain. Element #3 is different from the other six elements. But Elements #4, #5, #6 and #7 have very similar topologies. In fact, we have run many optimization tests and found that the optimized Element #3 is the best solution, which can not only deliver the prescribed wave properties, but also satisfy the structural requirements. We have even taken Elements #4, #5, #6 or #7 as the initial design configuration in the optimization, which however yielded no better elements than #3. This demonstrates that the configurations of Elements #4, #5, #6 and #7 have reached the limit, i.e., single configuration cannot realize all needed indices. All optimized elements except Element #1 are a combination of a space-coiling structure and cavities. As shown in Fig. 3B, the optimized elements can indeed display constant indices over the entire frequency range, showing apparent ultra-broadband non-dispersive property while maintaining relatively high transmission (Fig. 3C). We notice a low transmission region for Element #3 around 2.0 kHz, which is likely caused by a resonance associated with the relatively thin channels in the structure. Nevertheless, this low transmission only occurs within a very narrow frequency range [1870 Hz, 1940 Hz], and the average transmission of Element #3 remains high. In addition, the exact frequency corresponding to the minimal transmission (38.2%) of the optimized Element #3 is 1880 Hz, which is very close to the theoretically calculated FP resonance frequency of 1904 Hz. In this case, the optimized Element #3 indeed induces the FP resonance effect, which is not the case of the other six optimized elements (Supplementary Fig. S13b) because of their high transmissions over the frequency range. Simulated and measured acoustic fields in Fig. 3D clearly demonstrate the ultra-broadband frequency-independent performance in terms of energy focusing at the same focal depth. Moreover, the target range [1000 Hz, 3000 Hz] used in optimization is extended to [1000 Hz, 4000 Hz] as well, showing stable ultra-broadband properties. The observed focusing feature can also be maintained when the thermo-viscous loss is considered. In other words, thermal-viscous loss exerts limited effect, although the focusing efficiency inevitably becomes somewhat lower (Supplementary text S6). Consequently, the optimized asymmetrical topology can serve as an ideal candidate for ultra-broadband non-dispersive metasurface engineering.

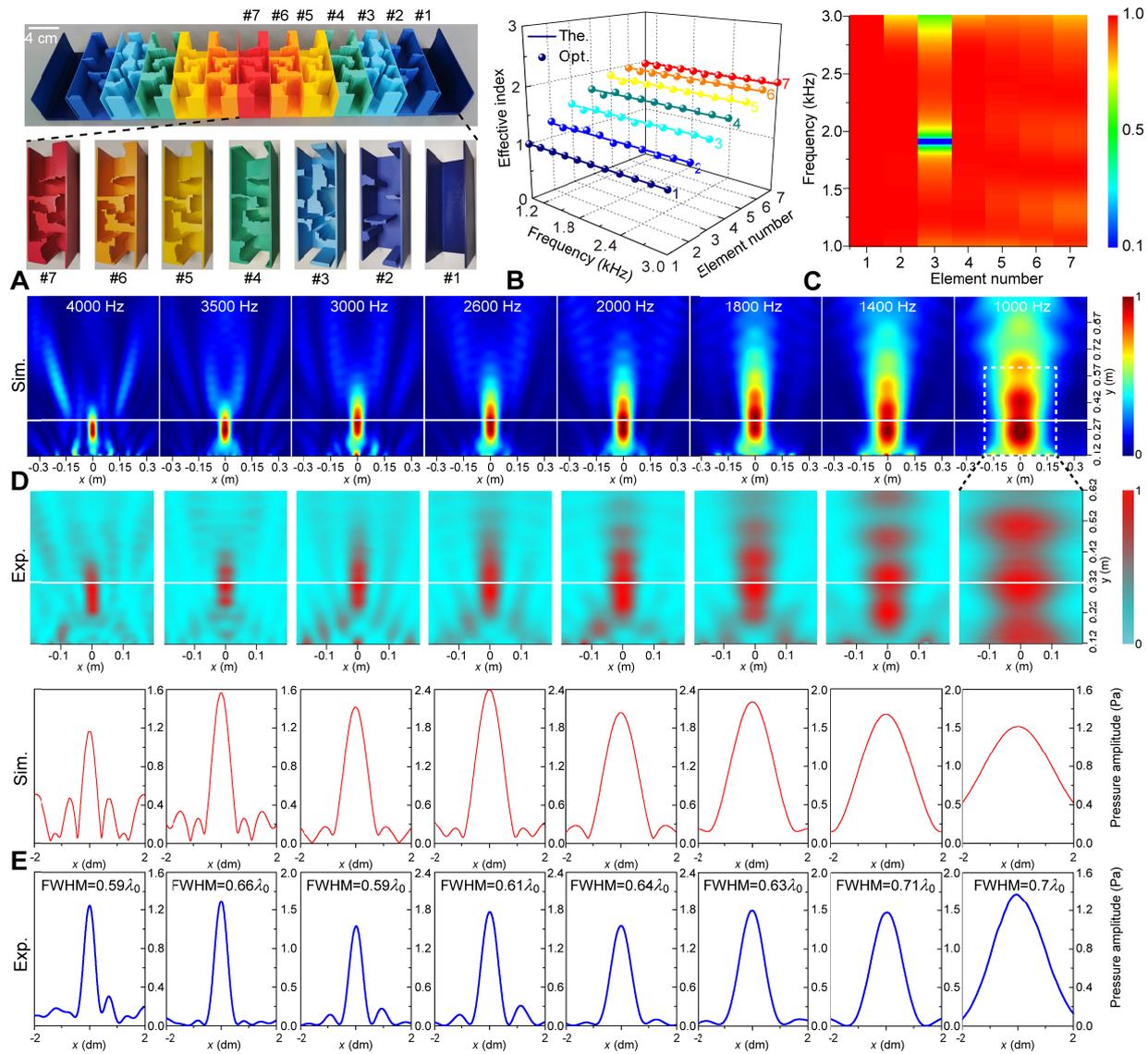

**Fig. 3. Ultra-broadband achromatic focusing using an inversely designed metasurface.** (**A**) A fabricated converging metasurface based on 13 optimized elements. (**B-C**) Theoretical (The.), optimized (Opt.) effective indices (**B**) and transmission coefficients (**C**) of all optimized elements within [1000 Hz, 3000 Hz]. (**D**) Simulated (Sim.) and measured (Exp.) acoustic amplitude fields at 8 representative frequencies within [1000 Hz,

4000 Hz]. The white solid line indicates the desired focal depth ($F_0$=0.2 m). The dashed line denotes the 0.4 m×0.5 m measured domain. For the same frequency, the simulated and measured fields use different normalization scales, which are the maximal pressure amplitudes. Results at different frequencies adopt different normalization scales. (**E**) Profiles of the simulated (Sim.) and measured focusing fields (Exp.) at the focal planes. The full widths at the half maximum (FWHM) of the measured focusing are shown as well, where $\lambda_0$ denotes the wavelength of the air.

**Engineering metasurface for ultra-broadband achromatic levitation**

In addition to beam steering and focusing, AMs can also generate more complex field patterns for other applications. For example, the realization of bottle beams (*37, 38*) allows stable single-sided acoustic levitation (*11*), which can trigger promising applications in the fields of analytical chemistry, biomedicine, biophysics and microassembly. So far, most demonstrations have been presented in narrow-bands. As illustrated in Fig. 4, we apply our inverse-design strategy to realize ultra-broadband stable single-sided acoustic levitation (Methods M1 and M4). A square-symmetrical metasurface is constructed to generate the bottle beam (Fig. 4A, Supplementary text S7). The optimized asymmetrical elements for levitation are also structurally more complex compared with the topologies shown in Figs. 2A and 3A (Supplementary text S7). The calculated required effective indices of the elements show that some elements are non-dispersive while the others are strongly dispersive (Fig. 4B). Figure 4C illustrates that the optimized indices of the elements are very close to the theoretical requirements. The non-dispersive elements have space-coiling geometries with local cavities, while the dispersive ones contain solid blocks and complex cavities. Because of the FP resonances involved (Supplementary Fig. S20), some optimized elements have low transmission in specific narrow-band ranges, which marginally affect the resulting functionality performances. Fortunately, the averaged transmission is guaranteed to be above 80% during the optimization process (Fig. 4D). It is found that only the optimized asymmetrical topologies with multiple scatterers can realize the complex non-dispersive and dispersive ultra-broadband phase manipulation while keeping a relatively high transmission. Figure 4E shows the generation of the beam within [16.5 kHz, 66 kHz], despite of a certain shift of the "dark" trapping regions. A small object can be suspended in mid-air under the pushing and pulling forces (*49*) originated from the scattering of these dark regions. In the measurements, we used a polyethylene ball of 0.022 g and observed its steady single-sided suspension in the air at representative frequencies (Fig. 4F, Method M7, Supplementary movies S1-S5). Limited by the single-frequency ability and fixed size of ultrasound transducers available on the market, we used the mainstream products and conducted the levitation tests for 32 kHz and 40 kHz. We also performed simulations by introducing thermo-viscous loss and confirmed that broadband performance is preserved (Supplementary text S8). Noted that most frequencies in Figs. 4E, 4F and S24 are not directly considered in the optimization. Therefore, it is remarkable that, the desired wave functionalities can still be realized over the entire bandwidth, as long as an appropriate number of discrete sampling frequency is selected (Supplementary text S26).

From structural design viewpoint, the optimization of metasurfaces is a classic multiple-solution problem. Although it is impossible to give the mathematical proof of the optimality, the finally achieved metasurfaces indeed show superior and expected functionalities. The optimized architecture satisfies the synthesized requirements in terms of phase matching, high averaged transmission, relatively wide air channels, one interconnected air region, maneuverable local sizes of solids and the fewest solid blocks, in addition to the prescribed wave functionalities.

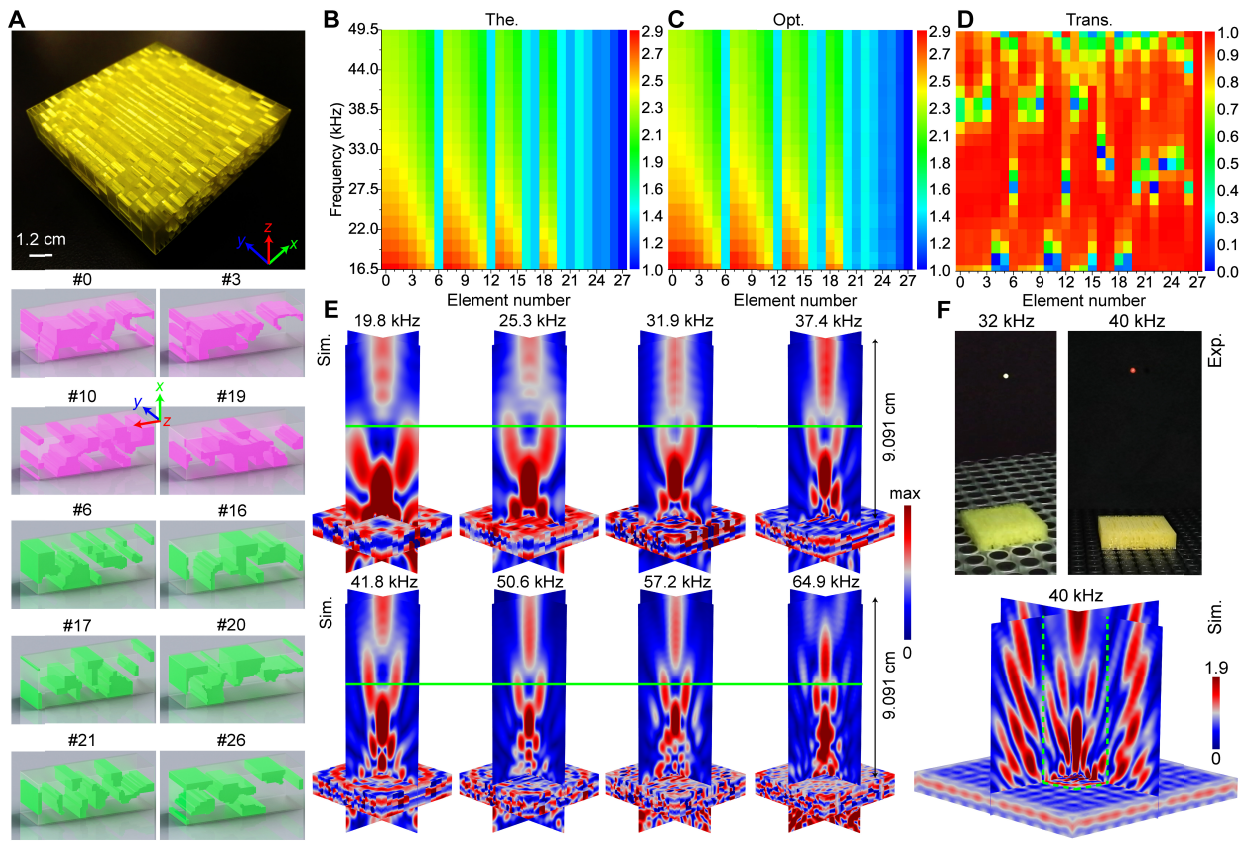

**Fig. 4. Ultra-broadband achromatic acoustic levitation using an inversely designed metasurface.** (**A**) A fabricated micron-scale 3D metasurface (5.46 cm×5.46 cm×1.2 cm) with 13×13 optimized elements (4.2 mm×4.2 mm×1.2 cm). The metasurface has square symmetry. Geometries of representative optimized elements showing novel topologies are presented below the metasurface. The pink and green graphs denote the representative non-linear dispersive and non-dispersive elements, respectively. (**B-D**) Theoretical (The.) (B) optimized (Opt.) (C) effective indices and transmission coefficients (Trans.) (D) of all optimized elements within [16.5 kHz, 49.5 kHz]. (**E**) Simulated (Sim.) bottle beam fields at 8 representative frequencies within [16.5 kHz, 66 kHz]. The green solid line indicates the predefined levitation location ($F_0$=4.545 cm). (**F**) Measured (Exp.) stable single-sided levitation in the air at two representative frequencies (32 kHz and 40 kHz) of a red polystyrene bead with a diameter of 3.4 mm. In the simulation (Sim.) mimicking the experiment, the metasurface is directly placed on the ultrasonic source array, inducing the desired bottle beam. The region inside the green dashed lines represents the area covered by the metasurface. The ultrasound levitation processes are presented in the Supplementary movies S1, S2, S4 and S5.

## Mechanism of ultra-broadband achromatic dispersions

To understand the underlying physics of the achieved ultra-broadband operations, we scrutinize the responses of the phase, effective impedance matrices and multiple scattering effects for representative metasurface elements (Fig. 5). Regardless of beam steering, focusing or levitation, the optimized asymmetrical space-coiling-cavity elements produce internal resonances in different regions of the structure at different frequencies (Figs. 5A-5C), thus compensating for the complex phase shift that results from the dispersion of individual components. Figures 5D-5F reveal that three representative elements also exhibit bi-anisotropy (*20*, *40*, *41*, Method M5). For a given functionality, the degree of bi-anisotropy of an element increases with its effective index or dispersion extent (Supplementary text S5). Consequently, more complex functionality usually requires stronger bi-anisotropy of the elements. As bi-anisotropy is related to the structural symmetries, stronger bi-anisotropy can enhance multiple-scattering and interaction of different individual components. We also discover a topological feature distinct from the traditional metasurfaces using space-coiling (*10*, *11*) and Helmholtz-resonator (*18*, *20*) structures, namely, multiple asymmetric solid blocks distributed in the space. To evaluate the interaction among the solid blocks, we obtain the transfer matrix profiles and the effective indices by dividing the element into different sections (Method M6, Supplementary text S2) and analyze the scattering properties. In this way, multiple scattering is evaluated (Figs. 5G-5I) and compared with the theoretical result. Regardless the way we divide the element, the single-scattering performance is totally different from the theoretical and retrieved ones, implying the existence of very strong multiple scattering, which plays a crucial role in determining the overall performance of the metasurface. Broadband properties would not have been possible without considering this effect. In addition, both the non-dispersive and dispersive elements exhibit similar multiple-scattering effects (Figs. S22 and S23), although the latter generally require stronger multiple scattering. For both non-dispersive and dispersive properties, every optimized element displays entirely different multiple scattering features (Figs. 5G-5I, Figs. S9, S10, S15, S16, S22, S23). Hence all the inversely designed metasurface elements exhibit diversified multiple scattering which can hardly be designed intuitively. On the other hand, topology optimization impliedly emphasizes the importance of the multiple scattering which can be regarded as a novel and additional degree of freedom of the design. Altogether, all optimized elements can collectively maintain suitable integrated internal resonances, bi-anisotropy and multiple-scattering effects simultaneously to realize the ultra-broadband functionality with high efficiency.

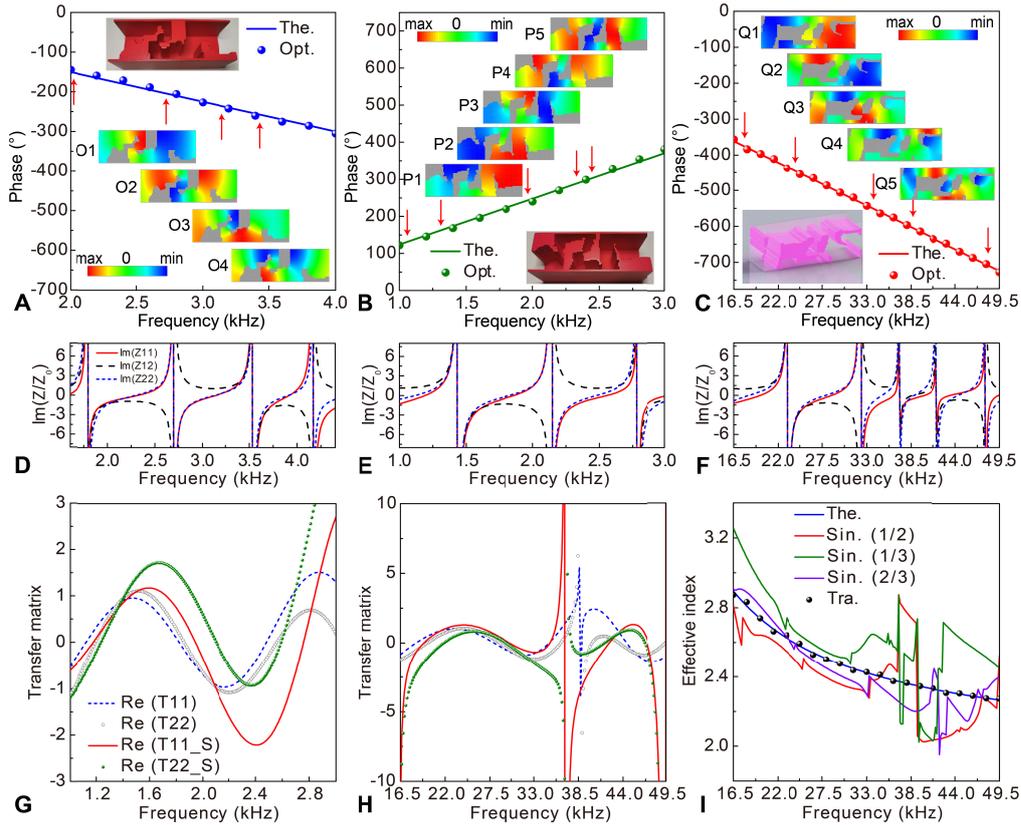

**Fig. 5. Ultra-broadband wave characterizations of representative optimized metasurface elements.** (**A-C**) Theoretical (The.) and optimized (Opt.) phase shift profiles of three representative optimized elements for wave steering (element #6) (A), focusing (element #7) (B) and levitation (element #0) (C) engineering. Insets display the layouts and special acoustic pressure fields of the optimized elements. O1-O4, P1-P5 and Q1-Q5 represent the fields at frequencies marked by red arrows from the left to right. These internal resonances collectively contribute to the rigorous phase compensations at different frequencies. (**D-F**) Impedance matrix profiles

(Method M5) of the optimized elements in A (D), B (E) and B (F). (**G-H**) Transfer matrix profiles of the elements in (A) and (C) without (T11, T22) or with (T11_S, T22_S) the single-scattering assumption (Method M6). (**I**) Comparison of the theoretical (Theo.), traditional retrieved (Tra.) and single-scattering (Sin.) induced effective index of optimized Element #0 for levitation engineering. The single-scattering results are deduced by the model divided at the half (1/2), one-third (1/3) and two-third (2/3) of Element #0 (Supplementary text S2).

## Discussion

By integrating the phase modulation, transmission, geometrical features and wave functionalities into design, we have methodically demonstrated ultra-broadband acoustic beam steering, focusing and levitation of achromatic metasurfaces by following a bottom-up topology optimization. To our knowledge, this represents a significant advance for designing metasurfaces with frequency-independent functionalities. With the thicknesses around 56.5–155.3%, 35.3–141.2% and 58.3–232.9% of the wavelengths within the frequency ranges [1600 Hz, 4400 Hz], [1000 Hz, 4000 Hz] and [16.5 kHz, 66 kHz], the inversely designed metasurfaces achieve remarkable relative bandwidths of 93.3%, 120% and 118.9% for three typical functionalities, respectively, validating the effectiveness of our approach. All optimized structures can represent a novel category of acoustic metasurfaces, which can support the asymmetric multiple scatterers, curve air channels and local cavities simultaneously. We discover that the optimized asymmetric metasurfaces can appropriately blend integrated internal resonances, bi-anisotropy and multi-scattering effects simultaneously, leading to the ultra-broadband inverse-customized dispersions. This high-efficiency ultra-broadband mechanism is expected to become a new and universal criterion for broadband AMs. In view of the robustness and universality of the proposed metasurfaces, the reported ultra-broadband topological features and customized dispersion mechanisms can provide new impetus to the whole field of optical (*42*, Supplementary text S10), acoustic and elastic metasurfaces. Since optimized elements inevitably cause the local zigzag boundaries, one may employ an explicit topology optimization (*43*) to create the smooth and parameterizable topological features. Alternatively, it is effective to consider the constraint of additive manufacturing in topology optimization to minimize the effect of manufacturing error.

It is possible to incorporate the multiple-scattering factor (MSF) as a critical index into the topology optimization in the future. However, due to the very complex features of the MSF at different frequencies for different dispersion properties, this is only possible if we can precisely control and optimize the effect of the multiple-scattering factor on the wave motions of a metasurface.

Limited by the computational capacity, the current optimization process is applied to a two-dimensional system. If 3D design is considered, the drastically increased design freedom will enable more complex wave functionalities with better performance and even wider bandwidth. In principle, if the micro/nano manufacturing and thermal-viscous loss can be solve effectively, MHz ultra-broadband achromatic ultra-sound devices of industrial value can also be inversely designed following the proposed procedure, if all dimensions of the metasurface are scaled down by 100 times. Finally, combing our designs with existing active control and deep learning techniques will make large-scale, digital, reconfigurable and integrated achromatic metasurface devices accessible.

## Materials and Methods

### M1. Topology optimization of a metasurface element

To construct a metasurface with desired functionality, we employ topology optimization to systematically design metasurface elements with the required phase distributions at different frequencies. Meanwhile we have to introduce a special physical constraint to ensure high transmission. During the optimization, an element may involve several enclosed regions, resulting in obvious narrow-band resonances. In addition, an optimized element usually has weak control ability of transmission phase if it cannot stay connected across the two ports of the design domain. Moreover, an optimized element cannot be effectively fabricated when its minimal solid component is too thin to guarantee the sufficient strength and manufacturing compatibility. Conceivably, very narrow air channels will result in the conspicuous viscous dissipation as well. Therefore, we should also introduce some geometrical constraints to make the optimized elements manufacturable and physically sensible. For any desired broadband functionality, the single-objective optimization with multiple constraints for a 2D rectangular design domain (Supplementary text S1) is formulated as

$$For: f_i \in [f_{\min}, f_{\max}] \ (i=1,2\cdots N_F), \tag{1}$$

$$Maximize: O(\Omega_D) = \frac{-max\left\{\sqrt{\frac{\sum_{i=1}^{N_F}(\phi_D^i)}{N_F}} \times \max_{\forall i \subset (1,2\cdots,N_F)}(\phi_D^i), \sqrt{\left[\frac{1}{N_F}\sum_{i=1}^{N_F}(\phi_D^i - \frac{\sum_{i=1}^{N_F}(\phi_D^i)}{N_F})^2\right]} \times \max_{\forall i \subset (1,2\cdots,N_F)}(\phi_D^i)\right\}}{360}, \tag{2}$$

$$\begin{aligned}Subject\ to: &\ \rho_i = 0\ or\ 1\ (i=1,2\cdots N_{EX} \times N_{EY}), \\ &\ \Omega_A \geq 1, \\ &\ \min_{\Omega_D}(\mathbf{w_A}) \geq w_{A0}, \\ &\ \min_{\Omega_D}(\mathbf{w_S}) \geq w_{S0}, \\ &\ T(\Omega_D) \geq 0.6, \end{aligned} \tag{3}$$

where $f_i$ denotes a sampling frequency of the target frequency range [$f_{\min}, f_{\max}$] which is uniformly divided into ($N_F$−1) frequency subintervals by $N_F$ discrete frequencies; $\Omega_D$ is the topological distribution within the 2D design domain (Supplementary Sec. S1); $O$ represents the objective function value for characterizing the violating extent of the theoretical phase shifts at all sampling frequencies; $\phi_D^i$ means the absolute difference between the desired phase shift and the theoretical one at a sampling frequency; $\rho_i$ indicates the material density of an element which can declare its attribute of the air (0) or solid (1); $N_{EX}$ and $N_{EY}$ are the numbers of the finite elements of the design domain along the *x* and *y* directions; $\Omega_A$ expresses the number of connected air domains containing at least one connected region covering from the two ports of the design domain; $\mathbf{w_A}$ and $\mathbf{w_S}$ stand for the arrays containing the sizes of all local air and solid components, respectively; $w_{A0}$ and $w_{S0}$ are the empirical values suggested by many optimization tests (*28*, *29*); $T$ represents the minimal value (beam steering) or the averaged value (focusing and levitation) of the transmission coefficients (Supplementary text S1) at $N_F$ sampling frequencies. Compared with broadband focusing and levitation, broadband beam steering requires relatively simpler control of dispersions. Apparently, stronger constraint of transmission can be utilized for the beam steering engineering. A constraint value of 0.6 is suggested by multiple optimization tests. In theory, the topology optimization of metasurfaces based on the specific phase and transmission properties is a classic non-convex, multiple-constraint, massive-variable optimization problem. For an given phase $\phi$, different structures with ($\phi \pm n \times 360°$, *n* is an integer) should deliver the same phase performance. Therefore, the optimization problem should have multiple-solutions, inferring that there should exist many microstructures with the same phase and transmission properties. In view of involving several strict geometrical constraints, our optimization model usually generate the similar asymmetric topologies after many runs.

Manuscript    Page **12** of **21**

To solve the inverse-design problem described by Eqs. (1)-(3), we select GA (*28-30*, *50*, Supplementary text S1) to perform topology optimization for metasurfaces to achieve the three customized functionalities. Note that the analysis on the effect of structural symmetry clearly shows that the asymmetrical topology of metasurface elements is the most beneficial feature for achieving perfect phase distributions and high transmissions (Fig. S5). The material parameters in optimization are chosen as: $\rho_{air}$=1.29 kg/m$^3$, $c_{air}$=340 m/s, $\rho_{solid}$=1230 kg/m$^3$ and $c_{solid}$=2230 m/s. Because the impedance of solids is much larger than that of the air, generally, the solids can be assumed to be acoustically hard to simplify the optimization (*28*). The target frequency ranges in optimization are predefined as [2000 Hz, 4000 Hz], [1000 Hz, 3000 Hz] and [16.5 kHz, 49.5 kHz] for the beam steering, focusing and levitation, respectively. In view of the different dispersive properties, the levitation engineering has to adopt more sampling frequencies ($N_F$=21) to capture the accurate phase shifts than those in the beam steering ($N_F$=11) and focusing ($N_F$=11) engineering (Supplementary text S9). To obtain structures with smooth edges in reasonable computing time, a "coarse to fine" two-stage strategy is applied in all optimizations (*28*, *29*). In the coarse stage, the design domain is meshed into 20×60 ($N_{EX}$=60, $N_{EY}$=20) pixels. In the fine stage, however, all optimized elements in the coarse stage serve as the initial population of new run of GA in the finer 40×120 ($N_{EX}$=120, $N_{EY}$=40) pixels. The parameters of minimal size constraints are set to $w_{A0}$=2mm, $w_{S0}$=1mm for the beam steering engineering, $w_{A0}$=2mm, $w_{S0}$=1mm for the focusing engineering, and $w_{A0}$=200 μm, $w_{S0}$=100 μm for the levitation engineering. The algorithm parameters of GA are the population size $N_p$=30, the crossover possibility $P_c$=0.9, the mutate possibility $P_m$=0.03, and the tournament competition group $N_{ts}$=21. To guarantee the near-optimality, every optimized element is selected from 5 optimizations with the same parameters. Numerous numerical tests show that sufficient generations can make the optimization converge to the ideal phase requirements. Specifically, the maximal generation number $E_G$ takes 10000 (coarse) and 10000 (fine) for the beam steering engineering, 3000 (coarse) and 10000 (fine) for the focusing engineering, and 2500 (coarse) and 5000 (fine) for the levitation engineering, respectively. It is worth noting that a more complex functionality contrarily demands fewer generations, implying easier capture of the evolution direction for a more complex phase requirement. All optimizations are conducted on a Linux cluster with Intel Xeon Platinum 8168 @ 2.70 GHz. The total time for an element optimization is about 93.3, 64.2 and 36.4 hours for the beam steering, focusing and levitation engineering, respectively. The performances of the phase shift and transmission coefficient in optimization are computed by the commercial finite element software ABAQUS 6.14-1.

## M2. Theory of designing ultra-broadband acoustic beam steering

Inspired by the generalized Snell's law (*1*), we can systematically construct a specific transverse phase gradient of the metasurface to realize a desired refractive beam for airborne sound. When waves are incident on an artificial structure, the refraction angle can be related to the incident angle and phase gradient (*1*)

$$\frac{1}{\lambda_t}\sin\theta_t - \frac{1}{\lambda_i}\sin\theta_i = \frac{1}{2\pi}\frac{d\phi}{dx}, \qquad (4)$$

where $\lambda_t$ and $\lambda_i$ indicate the wavelengths of two medium; $\theta_t$ and $\theta_i$ are the angles of refraction and incidence, respectively; and $d\phi/dx$ is the gradient of phase discontinuity along the interface between two media. Equation (4) clearly implies that only the design considering the variation of phase shift with frequency can accomplish the truly broadband wave manipulation. In other words, the designed structure for broadband exotic refraction of airborne sound has to keep the term $(c_{air}/\omega)\times(d\phi/dx)$ constant over the wide frequency range, where $\omega$ denotes the angular frequency. Here we perform topology optimization of 6 elements to construct the metasurface for broadband beam steering. The corresponding theoretical phase shift distributions for all elements are shown in Fig. S7. In addition, figure S11 depicts the theoretical simulations of beam steering using the ideal phase shift distributions.

## M3. Theory of designing ultra-broadband acoustic focusing

To focus the incident waves at a certain location away from the metasurface, the relative phase distribution provided by the metasuface should follow (*44*)

$$\phi(x,\omega) = \frac{\omega}{c_{air}}\left(\sqrt{x^2 + F_0^2} - F_0\right), \tag{5}$$

where $x$ and $F_0$ are the coordinate and focal length, respectively. To implement the frequency-independent metasurface, we have to make the transverse wavevector of $d\phi(x,\omega)/dx$ constant with any different given coordinate and frequency. Unlike the reported approaches for the optical achromatic metalens (*4-7*, *32*), we use topology optimization to directly design every element with the desired phase shifts at every sampling frequency, explicitly tailor-making all phase performances on demand. The theoretical phase shift distributions for all 7 elements are shown in Fig. S13. In addition, figure S17 depicts the theoretical simulations of focusing using the ideal phase shift distributions.

## M4. Theory of designing ultra-broadband acoustic levitation

To show the enormous potential of wavefront shaping, metasurfaces can be elaborately designed to generate the so-called bottle beam and then lead to the typical single-sided levitation (*11*, *37-39*). However, neither the existing ultrasonic phase arrays nor the two-layer metamaterial bricks are able to provide ideal phase manipulation only by a single structure without sophisticated regulation and equipment, not to mention the stable levitation over a broadband frequency range. To realize the levitation near the focal region, the superimposed phase distribution provided by the metasurface can be expressed as

$$\phi(x,y,\omega) = \frac{\omega}{c_{air}}\left(\sqrt{x^2 + y^2 + F_0^2} - F_0\right) + \phi_0(x,y),$$

$$\phi_0(x,y) = \begin{cases} 0, & (\sqrt{x^2+y^2} \le r_0) \\ 180, & (\sqrt{x^2+y^2} > r_0) \end{cases}, \tag{6}$$

where $x$ ($y$) and $F_0$ are the coordinate and the focal length created by the focusing decomposition field of the bottle beam, respectively; $\phi_0$ is the holographic signature phase; and $r_0$ means the region whose phase is opposite to the outer region. Apparently, a straightforward way to realize the broadband levitation is to make the spatial- and frequency-dependent transverse wavevector $d\phi(x, y, \omega)/dr$ constant for any location and frequency, where $r$ denotes the spatial coordinate $\sqrt{x^2 + y^2}$. To tactically design this complex functionality, we also use topology optimization to explore the metasurfaces with 13×13 elements possessing the phase distribution extremely close to the theoretical one (Fig. S20). Finally, a 3D inversely designed metasurface will be assembled by the 2D optimized 28 elements. In addition, figure S24 depicts the theoretical simulations of levitation using the ideal phase shift distributions.

## M5. Impedance matrix of a bi-anisotropic metasurface element

In recent years, bi-anisotropic metasurfaces have been shown to provide a new way for wave manipulations by realizing the independent control of the reflection and transmission phases or the difference in the reflection phases (*45-47*). In theory, asymmetrical element leads to the bi-anisotropic response at macroscale. In this case, the phase of reflection for both the forward and backward propagations is different (asymmetrical) (*20*), thus giving rise to the different scattering patterns depending on the direction of illumination. Since all optimized elements in our work are asymmetric, it is necessary to check the impedance matrices to evaluate their bi-anisotropic property. Using the four-microphone model (Fig. S1, *38*), we compute the responses in two cases. One case considers the nonreflecting boundaries at the right end of the model (#1), while, the other one adopts a hard wall at the right end (#2). The extracted pressure is expressed as $P_i^j$, where $i$ denotes the sequence number of the microphone and j represents the case number. The complex sound pressures at four probing positions for the two cases are respectively expressed as

$$\begin{bmatrix} -e^{-ik_0 y_1} & e^{-ik_0 y_1} \\ -e^{-ik_0 y_2} & e^{-ik_0 y_2} \end{bmatrix} \begin{bmatrix} A^1 & A^2 \\ B^1 & B^2 \end{bmatrix} = \begin{bmatrix} P_1^1 & P_1^2 \\ P_2^1 & P_2^2 \end{bmatrix}, \quad (7)$$

and

$$\begin{bmatrix} -e^{-ik_0 y_3} & e^{-ik_0 y_3} \\ -e^{-ik_0 y_4} & e^{-ik_0 y_4} \end{bmatrix} \begin{bmatrix} C^1 & C^2 \\ D^1 & D^2 \end{bmatrix} = \begin{bmatrix} P_3^1 & P_3^2 \\ P_4^1 & P_4^2 \end{bmatrix}, \quad (8)$$

where $k_0$ is the wave number.

With the metasurface element located at $y_0$, the pressures and velocities on the left (−) and right (+) sides of the element are respectively determined by

$$\begin{bmatrix} p^{-1} & p^{-2} \\ v^{-1} & v^{-2} \end{bmatrix} = \begin{bmatrix} e^{-ik_0 y_0} & e^{ik_0 y_0} \\ e^{-ik_0 y_0}/Z_0 & -e^{ik_0 y_0}/Z_0 \end{bmatrix} \begin{bmatrix} A^1 & A^2 \\ B^1 & B^2 \end{bmatrix}, \quad (9)$$

and

$$\begin{bmatrix} p^{+1} & p^{+2} \\ v^{+1} & v^{+2} \end{bmatrix} = \begin{bmatrix} e^{-ik_0 (y_0+d)} & e^{ik_0 (y_0+d)} \\ e^{-ik_0 (y_0+d)}/Z_0 & -e^{ik_0 (y_0+d)}/Z_0 \end{bmatrix} \begin{bmatrix} C^1 & C^2 \\ D^1 & D^2 \end{bmatrix}, \quad (10)$$

where $p$, $v$, $d$ and $Z_0$ are the pressure, the particle velocity, the width of metasurface element and impedance of the air, respectively.

Accordingly, the transfer matrix **T** relating the pressures and velocities on two sides becomes

$$\begin{bmatrix} T_{11} & T_{12} \\ T_{21} & T_{22} \end{bmatrix} = \begin{bmatrix} p^{+1} & p^{+2} \\ v^{+1} & v^{+2} \end{bmatrix} \begin{bmatrix} p^{-1} & p^{-2} \\ v^{-1} & v^{-2} \end{bmatrix}^{-1}. \quad (11)$$

Therefore, the impedance matrix **Z** can be calculated by

$$\begin{bmatrix} Z_{11} & Z_{12} \\ Z_{21} & Z_{22} \end{bmatrix} = \begin{bmatrix} -\dfrac{T_{22}}{T_{21}} & -\dfrac{1}{T_{21}} \\ \dfrac{T_{12}T_{21} - T_{11}T_{22}}{T_{21}} & -\dfrac{T_{11}}{T_{21}} \end{bmatrix}. \quad (12)$$

Furthermore, the scattering matrix **S** can be obtained by

$$\begin{bmatrix} r^+ & t^- \\ t^+ & r^- \end{bmatrix} = \begin{bmatrix} \dfrac{(Z_{11}-Z_0)(Z_{22}+Z_0)-Z_{21}Z_{12}}{(Z_{11}+Z_0)(Z_{22}+Z_0)-Z_{21}Z_{12}} & \dfrac{2Z_{12}Z_0}{(Z_{11}+Z_0)(Z_{22}+Z_0)-Z_{21}Z_{12}} \\ \dfrac{2Z_{21}Z_0}{(Z_{11}+Z_0)(Z_{22}+Z_0)-Z_{21}Z_{12}} & \dfrac{(Z_{11}+Z_0)(Z_{22}-Z_0)-Z_{21}Z_{12}}{(Z_{11}+Z_0)(Z_{22}+Z_0)-Z_{21}Z_{12}} \end{bmatrix}, \quad (13)$$

where $t^+$ and $t^-$ mean the forward and backward transmission coefficients; and $r^+$ and $r^-$ are the forward and backward reflection coefficients, respectively.

### M6. Multiple-scattering factor of an optimized element

In the field of metasurfaces, the use of multiple scattering effects to modulate the dispersions and effective index has never been considered because of the complexity and harsh demand in the microstructural design. Since most elements in this paper have many distributed solid blocks, it becomes extremely complicated to regard every block as an independent scatterer. Here we adopt two steps to analyze the variation of the effective index to investigate the existence of multiple scattering and characterize the multiple-scattering effects of the optimized elements. In the first step, the traditional inverse technique (*49*) is utilized to retrieve the corresponding transfer matrix **T_O**, scattering matrix **S_O** and effective index $n_O$ of an optimized element. In the second step, the optimized element is divided into several substructures (Supplementary text S2) whose respective transfer matrix **T**$_i$ ($i$=1, 2···$m$) is calculated respectively, where $m$ is the total number of substructures. Then we can get the effective transfer matrix **T_S**=**T**$_m$·**T**$_{m-1}$···**T**$_2$·**T**$_1$ of the element based on the single scattering assumption. As a result, the effective single-scattering index $n_S$ will be induced by the effective single-scattering transfer matrix **T_S**. To characterize the multiple-scattering extent of the element, we define the multiple-scattering factor $\chi$ as

$$\chi = 2\frac{|n_O - n_S|}{n_O + n_S}, \qquad (14)$$

where both $n_O$ and $n_S$ are positive for all optimized elements within the target frequency range. Note that $\chi$ becomes zero for the traditional space-coiling (*10*, *11*, *14*, *24*) and Helmholtz-resonator metasurfaces (*12*, *16-20*, *23*). However, the obvious multiple scattering effects exist in our optimized elements.

**M7. Simulations and experiments of ultra-broadband wave functionalities**

To verify the desired ultra-broadband beam steering, focusing and levitation, we investigate the full-wave propagation through assembling inversely designed metasufaces by using the acoustics module in the commercial software COMSOL Multiphysics 5.3. The background medium is air and the solid metasurfaces are treated as rigid material, so the acoustically-hard boundaries are used on the surface. Plane wave radiation boundary condition is adopted on the outer boundaries of the computational domain to eliminate the reflected waves, and the plane wave is excited on the incident port. To consider the thermal-viscous loss in the simulations, the thermoacoustics interface is used to compute the variations of the acoustic pressure, velocity, and temperature. The interface is required to accurately calculate the acoustic field in geometries of small dimensions, so the thermoacoustics domain is introduced in the region with element-cells containing complex acoustic channels. We then introduce the acoustic-thermoacoustic boundary nodes to couple the thermoacoustic domain to the acoustic domain. Viscosity and thermal conduction near the hard walls become important because they create a viscous and thermal boundary layer where losses are significant. Therefore, the boundary layers are meshed with a dense element distribution in the normal direction along specific boundaries. The plane wave radiation boundary condition is also adopted on the outer boundaries of the simulation domains.

In the acoustic experiments on beam steering and focusing, we used the same platform (Fig. S17e) to measure the transmitted acoustic fields. A loudspeaker array, located 0.15 m away from the input interface of the metasurface surrounded by sound-absorbing sponges, was used as the incident plane waves; while a mounted microphone was connected to the B&K device (Type 3160-A-042) to measure the acoustic pressure by moving in the 40cm×50cm scanning area. The measured signals at each position were averaged over four measurements to reduce the effect of noise. Using the Fourier transform, the whole acoustic filed was obtained after the scanning measurement.

In the ultrasonic experiments on levitation, we use a self-developed ultrasound array system (Fig. S25) to generate the ultrasonic plane waves, impinging on and propagating through the inversely designed metasurface. The amplitude and the phase of every ultrasonic transducer was set to be the same as before for plane wave generation. Since the metasurface can cover about 5×5 elements, only 12×12 or 10×10 ultrasonic transducers were open for the present ultrasonic experiments with the voltage of 20.3 V. Upon deploying the ultrasound array system, we put a polystyrene bead around the desired levitation position. As a result, the polystyrene bead was suspended in mid-air by the ultrasonic radial force. To characterize the ultrasonic field, the mounted microphone connected the B&K device (Type 3160-A-042) measured the acoustic pressure by moving within a 3D scanning area. The signals at each position were obtained out of four measurements. Using the Fourier transform, the whole acoustic filed was obtained after the scanning measurement.

**M8. Sample fabrication**

For the ultra-broadband beam steering and focusing, a 3D-printing machine, UnionTech Lite800HD (Company: Wenext) with a printing resolution of 200 μm, was used to fabricate the metsauface samples made of low-viscosity photosensitive resin with a mass density of 1250 kg/m$^3$ and an elastic modulus of 2370-2650 MPa.

As for the ultra-broadband levitation, the metasurface sample was 3D-printed by a commercial machine, Micro Scale 3D Printing System nanoArch® S140 (Company: BMF Material Technology Inc.) with a printing resolution being 10 μm. The printing materials was a low-viscosity high-temperature resistant resin whose mass density and elastic modulus are 1150 kg/m$^3$ and 4.2 GPa, respectively.

**Acknowledgments**

**Funding:** This work was supported by the Hong Kong Scholars Program (No. XJ2018041), the National Natural Science Foundation of China (Grant Nos. 11802012, 11991031, 11902171 and 11534013), the Postdoctoral Science Foundation (2017M620607), the Fundamental Research Funds for the Central Universities (Grant No. FRF-TP-17-070A1), the Sino-German Joint Research Program (Grant No. 1355), the German Research Foundation (DFG, Project No. ZH 15/27-1), an Emerging Frontiers in Research and Innovation grant from the National Science Foundation (Grant No. 1641084) and the Shenzhen key laboratory of ultrasound imaging and therapy (ZDSYS201802061806314). **Author contributions:** H. W. D and C. S conceived the origin idea. S. A. C, Y. S. W and L. C supervised the project. H. W. D formulated and accomplished all optimizations and designs of metasurfaces. H. W. D, S. D. Z and C. S conducted the physical analyses and simulations of wave functionalities. S. D. Z performed the experiments of ultra-broadband beam steering and focusing. W. Q developed the integrated system of ultrasound transducers. H. W. D, J. Z and W. Q carried out the ultrasound experiments of ultra-broadband levitation. H. W. D and C. S wrote the original manuscript. All authors discussed the results and commented to the manuscript. S. A. C, Y. S.W, L. C and C. Z revised the manuscript. Authors like to thank Prof. Bilong Liu (Qingdao University of Technology, PR China), Dr. Junfei Li (Duke University, USA), Mr. Xuan-Bo Miao (Tianjin University, PR China) and Mr. Hong-Tao Zhou (Tianjin University, PR China) and for their helpful participations. **Competing interests:** Authors declare no competing interests. **Data and materials availability:** All data needed to evaluate the conclusions in the paper are present in the paper or the supplementary materials.

**Supplementary Materials**

Movie S3. Response of the same polystyrene bread putting over the surface of the same ultrasonic source, when the metasuaface is absent.
Movie S4. Oblique view of the very stable ultrasonic levitation at 32 kHz using the inverse-designed 3D ultra-broadband metasuraface.
Movie S5. Front view of the very stable ultrasonic levitation at 32 kHz using the inverse-designed 3D ultra-broadband metasuraface.
Movie S6. Evolution of optimized Element #2 for the ultra-broadband beam steering from a random initial design.
Movie S7. Evolution of optimized Element #4 for the ultra-broadband beam steering from an "Air" design.
Movie S8. Evolution of optimized Element #5 for the ultra-broadband beam steering from an "Air" design.